# Preprint version of the manuscript

Title:

**Time-efficient combined morphologic and quantitative joint MRI based on clinical image contrasts - An exploratory *in-situ* study of standardized cartilage defects**

Authors:

Teresa Lemainque[1], Nicola Pridöhl[1], Shuo Zhang[2], Marc Huppertz[1], Manuel Post[1], Can Yüksel[1], Masami Yoneyama[3], Andreas Prescher[4], Christiane Kuhl[1], Daniel Truhn[1], Sven Nebelung[1]

Institutions:

[1] Department of Diagnostic and Interventional Radiology, Medical Faculty, RWTH Aachen University, 52074 Aachen, Germany

[2] Philips GmbH Market DACH, Hamburg, Germany

[3] Philips Japan, Tokyo, Japan

[4] Institute of Molecular and Cellular Anatomy, RWTH Aachen University, 52074, Aachen, Germany

Corresponding author:

Teresa Lemainque, PhD
Klinik für Diagnostische und Interventionelle Radiologie
Uniklinik RWTH Aachen
Pauwelsstr. 30
52074 Aachen
Germany
E-Mail: tlemainque@ukaachen.de
Tel: +49 241 80 37843

**Funding declaration:**

This work was supported by the START program of the Faculty of Medicine of the RWTH Aachen University (No. 692316). DT is supported by the European Union's Horizon Europe programme (ODELIA, 101057091), by grants from the Deutsche Forschungsgemeinschaft (DFG) (TR 1700/7-1), and the German Federal Ministry of Education and Research (SWAG, 01KD2215A; TRANSFORM LIVER, 031L0312A). SN is funded by grants from the Deutsche Forschungsgemeinschaft (DFG) (NE 2136/3-1).

# Abstract

**Objectives:**

Quantitative MRI techniques such as T2 and T1ρ mapping are beneficial in evaluating cartilage and meniscus. We aimed to evaluate the MIXTURE (**M**ulti-**I**nterleaved **X**-prepared **T**urbo-Spin Echo with Int**U**itive **RE**laxometry) sequences that provide morphologic images with clinical turbo spin-echo (TSE) contrasts and additional parameter maps versus reference TSE sequences in an in-situ model of human cartilage defects.

**Materials and Methods:**

Prospectively, standardized cartilage defects of 8mm, 5mm, and 3mm diameter were created in the lateral femora of 10 human cadaveric knee specimens (81±10 years, nine male/one female). Using a clinical 3T MRI scanner and knee coil, MIXTURE sequences combining (i) proton-density weighted fat-saturated (PD-w FS) images and T2 maps and (ii) T1-weighted images and T1ρ maps were acquired before and after defect creation, alongside the corresponding 2D TSE and 3D TSE reference sequences. Defect delineability, bone texture, and cartilage relaxation times were quantified. Inter-sequence comparisons were made using appropriate parametric and non-parametric tests.

**Results:**

Overall, defect delineability and texture features were not significantly different between the MIXTURE and reference sequences. After defect creation, relaxation times increased significantly in the central femur (for T2) and all regions combined (for T1ρ).

**Conclusion:**

MIXTURE sequences permit time-efficient simultaneous morphologic and quantitative joint assessment based on clinical image contrasts. While providing T2 or T1ρ maps in clinically feasible scan time, morphologic image features, i.e., cartilage defect delineability and bone texture, were comparable between MIXTURE and corresponding reference sequences.

**Clinical relevance statement:** (max. 40 words)

Equally time-efficient and versatile, the MIXTURE sequence platform combines morphologic imaging using familiar contrasts and excellent image correspondence (versus corresponding reference sequences) with quantitative mapping techniques, thereby increasing the diagnostic information of routine scan protocols beyond mere morphology.

## Key points

- Combined morphologic and quantitative MIXTURE sequences based on 3D TSE contrasts were studied in an in-situ human cartilage defect model.
- Morphologic image features, i.e., defect delineability and bone texture, were similar between the MIXTURE and the reference sequences.
- MIXTURE allows time-efficient simultaneous morphologic and quantitative knee joint assessment based on familiar clinical image contrasts.

## Keywords

Magnetic resonance imaging, osteoarthritis, cartilage, knee joint

# 1. Introduction

Osteoarthritis is a chronic joint disease with increasing prevalence due to aging and obesity [1]. Clinically, MRI is well-suited for diagnosing cartilage degeneration as the hallmark change of osteoarthritis. Traditional MRI sequences such as proton density (PD)-weighted (-w) fat-saturated (FS) sequences focus on cartilage morphology, i.e., surface integrity and intra-tissue signal [2]. They are insensitive to early degenerative changes of tissue morphology, such as partial-thickness defects and fibrillation [3] [4].

Quantitative MRI techniques such as T2 or T1ρ mapping may be beneficial in detecting such changes – at a potentially reversible stage [5] [6]. Consensus prevails that adding T2 maps to the routine knee protocol improved sensitivity in detecting cartilage lesions significantly [3] [6]. The literature is less clear on T1ρ: Even though the association between cartilage degeneration and prolongation of T1ρ relaxation is well established [7] [8], its potential diagnostic benefits remain to be ascertained. Both mapping techniques have in common that their more widespread adoption is hampered by prohibitively long scan times and other challenges [9] [10].

Combined morphologic and quantitative sequences, such as quantitative double-echo in steady-state (qDESS), provide morphologic images and T2 maps in clinically feasible scan times [11] [12] [13]. QDESS sequences are diagnostically equivalent to conventional clinical MRI protocols; therefore, they are theorized to (partially) substitute the routine knee protocol while providing additional T2 maps [14]. Nevertheless, the morphologic images lack the clinically familiar contrasts of state-of-the-art turbo-spin-echo (TSE) sequences. The **M**ulti-**I**nterleaved **X**-prepared **T**urbo-Spin Echo with Int**U**itive **RE**laxometry (MIXTURE) sequence provides an alternative platform for combined imaging [15]. These sequences are designed to acquire at least two morphologic images of different contrast weightings using variable pre-pulses. Because pre-pulses, echo times, and spin-lock durations are freely adjustable, the sequences provide quantitative T2 or T1ρ maps as a ‘by-product’ of the morphologic images. In contrast to qDESS, MIXTURE is based on a 3D TSE acquisition. Preliminary studies have explored its principal clinical applicability, but systematic comparisons with reference sequences and standardized pathologies are lacking [16] [17] [18] [19] [20].

This study aimed to evaluate MIXTURE sequences in two principal configurations for their clinical usage and against corresponding 2D and 3D TSE reference sequences in an in-situ model of standardized cartilage defects. We hypothesized that the morphologic MIXTURE images are diagnostically on par with their reference sequence counterparts while additionally providing T2 and T1ρ maps in clinically feasible time frames.

# 2. Materials and methods

## 2.1 Study design

The local Institutional Review Board approved this prospective in-situ imaging study on human cadaveric knee joint specimens (Ethical Committee, RWTH Aachen University, EK180/16) conducted in 2022 and 2023. Fresh-frozen and non-fixated knee joint specimens from body donors who had given written informed consent prior to study initiation were provided by the local Institute of Anatomy (RWTH Aachen University). Moderate-to-severe cartilage degeneration of the lateral compartment, such as substantial tissue loss or focal lesions, was screened for during standard clinical scanning (using 2D TSE PD-w FS imaging) and defined as an exclusion criterion. Based on a preliminary analysis of the first three specimens, a minimum sample size of 8 was calculated using a statistical power of 80%, a significance level of 0.01, and an effect size (i.e., Cohen's d [21]) of 1.24. Hence, ten knee joint specimens were included.

## 2.2 Workflow

Specimens were left to thaw at room temperature for 24 hours. MR imaging was performed twice, i.e., before and after creating standardized cartilage defects. On day 1, pre-defect MR imaging was performed. The specimens were kept at 5°C overnight. On day 2, cartilage defects were created, and post-defect MR imaging was performed immediately afterward.

## 2.3 Cartilage defects

**Figure 1** presents the standardized step-wise creation of the cartilage defects. First, the knee joint was accessed through a median longitudinal skin incision and a medial peri-patellar approach. Once the joint was flexed, the patella was everted laterally to fully expose the joint. Second, the weight-bearing region of the lateral femoral condyle was identified. Three 3 mm, 5 mm, and 8 mm diameter defects were created in the lateral femoral condyle perpendicular to the condyle's bone contour. The cartilage tissue was removed using skin biopsy punches of corresponding diameters and surgical scalpels. Particular care was taken to maintain the integrity of the subchondral lamella. Third, the joint was thoroughly and continuously irrigated with 0.9% saline solution to remove surgical debris and excess air. Fourth, the joint was sutured layer-wise.

## 2.4 MR image acquisition

All scans were performed on a 3.0 T MRI scanner (Elition X, Philips, Best, The Netherlands) using an eight-channel transmit-receive knee coil. Specimens were positioned feet-first, supine, and in approximately 30° of flexion in line with clinical positioning. Two MIXTURE sequences were acquired as (i) PD-w FS images with T2 maps (scan time 4:59 min) and (ii) T1-w images with T1ρ maps (scan time 6:30 min) followed by the respective weightings' reference 2D TSE and 3D TSE sequences. MIXTURE sequence (i) acquired a PD-w FS image (using a Spectral Attenuated Inversion Recovery [SPAIR] pre-pulse) and a T2-weighted morphologic image (using a T2-preparation of 50 ms) in an interleaved manner. Voxel-wise T2 relaxation times were determined based on the two images by monoexponential fitting, and T2 maps were subsequently reconstructed on the scanner workstation using prototype software. MIXTURE sequence (ii) acquired a T1-w (without preparation) and two spin lock-prepared T1ρ-w FS morphologic images (using T1ρ-preparations of 25 ms and 50 ms, respectively, and SPAIR pre-pulses). Voxel-wise, T1ρ relaxation times were determined based on the three images by mono-exponential fitting, and T1ρ maps were reconstructed accordingly. Further details may be found in the literature [15]. 2D TSE reference PD-w FS and T1-w sequences were included per our clinical knee protocol. 3D TSE reference PD-w FS and T1-w sequences were obtained from the vendor and included. **Table 1** summarizes the sequence parameters. Notably, as 3D TSE acquisitions, the reference 3D TSE and MIXTURE sequences can, in principle, be acquired at isotropic resolution. In this study, however, we aimed to match the 3D TSE sequences to the 2D reference TSE sequences, i.e., the clinical reference standard, for voxel-to-voxel comparisons. Consequently, the 3D TSE sequences were acquired analogously to the 2D TSE sequence, i.e., using thicker slices and higher in-plane resolution than achieveable with isotropic image acquisitions.

## 2.5 Image analysis

Quantitative analyses were performed in Python (version 3.9.9) [22].

### 2.5.1 Defect delineability

Cartilage defect delineability was assessed on the PD-w FS sequences using line profiles manually annotated in ITK-SNAP (v3.8) [23] [24] by N.P. (pre-graduation medical student, two years of experience in medical imaging) and visually verified by S.N. (board-certified musculoskeletal radiologist, ten years of experience). Line profiles were placed through the defect and adjacent cartilage on the sagittal post-defect PD-w FS image that centrally bisected the defects (**Figure 2a**). As projections of the signal intensity (SI) along their course, SI line profiles were extracted from the 2D TSE, 3D TSE, and MIXTURE PD-w FS images and normalized to the maximum of 1 (**Figure 2b**). For every SI line profile, full width at half maximum (FWHM, **Figure 2c**) and edge width (EW, **Figure 2d**) were evaluated as surrogates of defect delineability. More specifically, a parallel line was defined at half maximum between the background signal level of cartilage and the maximum SI along the line profile. The horizontal distance between the intersections of this line with the SI line profile was determined as the defect's FWHM. Similarly, two vertical lines per defect shoulder defined the 10% and 90% maximum SI. The horizontal distance between these two lines was determined as the respective defect shoulder width, and EW was calculated as the mean of both defect shoulder widths.

### 2.5.2 Bone texture features

The bone texture on the T1-w sequences was quantified using radiomic features (**Figure 3**). In ITK-SNAP, circular regions of interest (ROI) with a diameter of 40 pixels were defined directly adjacent to the 5 mm defect (**Figure 3a**) on the same sagittal post-defect slice as above. Before computing texture features, the stacks of the 2D TSE, 3D TSE, and MIXTURE T1-w images were normalized between the SI values 0 and 1 (**Figure 3b**). Guided by earlier studies [25], we focused on variance, (joint) energy, (joint) entropy, and inverse difference (synonymous with "homogeneity1"[PyRadiomics]) to quantify the spatial distribution of SI values, characterize the underlying bone structure, and capture what the radiologist assesses on the microstructural level (**Figure 3c**). The texture features were determined using PyRadiomics [26]. **Variance** is a first-order feature that measures SI value spread within the ROI; high variance indicates high heterogeneity and large differences from their mean SI. Entropy, energy, and inverse difference are gray level co-occurrence matrix features. The gray level co-occurrence matrix quantifies how often different neighboring voxel value pairs are present within the ROI. **Entropy** measures disorder or complexity; high entropy indicates bone tissue with a complex texture characterized by diversely varying neighboring SI values. **Energy** measures textural uniformity; high energy indicates many repetitions of the same neighboring SI values. **Homogeneity** measures local image uniformity; high homogeneity indicates more uniform gray levels. High entropy, energy, and inverse difference values indicate more randomness, homogeneous patterns, and local homogeneity [26] [27].

### 2.5.3 Quantitative parameter maps

N.P. segmented the femoral and tibial cartilage plates on the MIXTURE PD-w FS images using ITK-SNAP. The central bisecting slice through the defects (post-defect) and the corresponding original slice (pre-defect) were segmented. The femoral cartilage was divided into an anterior ('aF'), central ('cF'), and posterior region ('pF') based on the outer contours of the lateral meniscus' anterior and posterior horns. The tibial cartilage ('T') was segmented as one region. All segmentation outlines were reviewed and adjusted by T.N. and S.N.. T2 and T1ρ values were computed (pre-defect and post-defect) and provided as mean ± standard deviation for each region and the entire lateral femorotibial compartment.

## 2.6 Statistical analysis

Statistical analysis was performed by N.P.,T.N., and S.N. using Graph Pad Prism (v9.5.1, San Diego, CA). Inter-sequence comparisons of FWHM, EW, and radiomic texture features were performed using repeated measures ANOVA followed by Tukey-Kramer post-hoc test. Pre- and post-defect T2 and T1ρ relaxation times were comparatively evaluated per region and overall using Wilcoxon matched-pairs signed-rank tests. To reduce the number of statistically significant but clinically likely irrelevant findings, the family-wise significance level was set to $\propto=0.01$. Multiplicity-adjusted p-values are provided.

## 3. Results

*Study cohort*

Ten knee joint specimens (age 81.1±10.4 years [mean±standard deviation]; range 68–96 years; 9/1 male/female) were included. Standardized cartilage defects were successfully created in all specimens.

*Qualitative evaluation*

In PD-w FS images, the cartilage defects were clearly discernable, and the cartilage tissue had the characteristic layer-wise configuration and intermediate SI in all sequences. Menisci and bone marrow appeared homogeneously dark, i.e., suppressed, while intraarticular fluid was homogenously bright (**Figure 4**). In the T1-w images, the macro- and microstructural bone texture appeared slightly less blurry in the 2D TSE sequence, particularly compared to the MIXTURE image (**Figure 5**). Contrast and noise levels appeared largely similar.

*Quantitative evaluation of defect delineability*

**Table 2** presents the metrics of defect delineability, i.e., FWHM and EW values. FWHM values were substantially lower than the nominal defect diameters but overall largely similar between the sequences. For the 5 mm defects, however, the 2D TSE sequence yielded significantly higher FWHM values than the MIXTURE sequence ($p=0.005$). On average, EW values were lower for the 2D TSE than for the 3D TSE and MIXTURE sequences, even though statistically not significant. The latter two sequences exhibited largely similar EW values.

*Quantitative evaluation of bone texture*

Voxel SIs contained in the ROI were spread out along 43±10 (2D TSE), 49±9 (3D TSE), and 42±9 (MIXTURE) bins, indicating a comparable spread of voxel SI distributions. The radiomic feature analysis indicated comparable bone texture feature values between the sequences (**Figure 6**). When comparatively evaluating the individual features, significant differences were only found between the 3D TSE and MIXTURE sequences with significantly higher energy and homogeneity values (and significantly lower entropy values) determined for MIXTURE vs. 3D TSE.

*Quantitative parameter maps*

We observed increased T2 and T1ρ relaxation times after defect creation for all studied regions (**Table 3**). These increases were mainly non-significant except for T2 in the central femur, where the defects were located (pre-defect, 51±4 ms; post-defect, 56±4 ms; $p=0.002$), and for T1ρ when considering all regions together (pre-defect, 40±4 ms; post-defect, 43±4 ms; $p=0.004$).

## 4. Discussion

Our study evaluated the image quality of MIXTURE PD-w FS and T1-w sequences relative to corresponding 2D and 3D TSE reference sequences. Focusing on the delineability of cartilage defects and quantitative bone texture features, we found that MIXTURE sequences were largely equivalent regarding image contrast, morphologic correspondence and coherence, and quantitative features. Simultaneously, MIXTURE sequences provided quantitative T2 or T1ρ maps with little additional scan time. Thereby, MIXTURE sequences increase the diagnostic information of routine scan protocols beyond mere morphology and may complement (or in parts even replace) current knee MRI protocols.

The primary advantage of MIXTURE sequences is their TSE-derived image contrast. Since their introduction to the clinic in the early 1990s, TSE sequences have been considered the standard for knee MRI; thus, radiologists are used to these images, and the American College of Radiology even formally recommends their usage [28]. MIXTURE sequences obviate the need for radiologists to familiarize themselves with other contrasts. Additionally, the sequence architecture is flexible and may be adjusted to other TSE-based weightings with or without fat saturation. A broad spectrum of sequence combinations can thus be efficiently acquired at each institution's discretion.

Specifically, we evaluated a PD-w FS sequence with T2 maps and a T1-w sequence with T1ρ maps acquired with 43 slices across the joint in 5 and 6.5 min, respectively. Previously, Kijowski et al. highlighted the clinical potential of adding T2 maps to the routine protocol [3]. Even though the diagnostic benefit of T1ρ maps remains unclear, adding more quantitative images to the morphologic standard images seems well-justified. MIXTURE needs at least two morphologic images, which require more acquisition time than a single image.

The MIXTURE PD-w FS sequence depicted the cartilage defects with a level of contrast and sharpness similar to the reference sequences. By trend, EW and FWHM values of the 2D TSE sequences were lower and closer to the nominal defect diameters, respectively, than those of the corresponding MIXTURE and 3D TSE sequences. This finding indicates slightly less clear defect delineability of the latter sequences and may be due to the higher echo train lengths [29] [30] or the choice of the refocusing pattern that, besides affecting image contrast and SNR, also influences image blurring [31].

Increased blurring, likely secondary to the choice of the refocusing pattern, was observed for the MIXTURE T1-w images and confirmed by the radiomic analysis of bone texture. Bone texture was significantly more homogeneous in the MIXTURE T1-w sequence, which may translate into a loss of micro-textural detail with as-yet unknown clinical relevance.

When designing the study, we aimed to compare cartilage and bone texture voxel-wise. To this end, we matched the image resolutions of all sequences, both in plane and through plane. Yet, this approach precluded the possibility of performing multiplanar reconstructions, an inherent feature of isotropic 3D sequences, which is a prerequisite for precise tissue segmentation (of cartilage and meniscus) for the analysis of morphometry and relaxivity [32] [33].

Quantitative analyses indicated increased post-defect relaxation times. Surgical tissue damage and fluid pressurization during preparation likely caused tissue swelling and edema. While the exact compositional and structural correlates of prolonged T1ρ and T2 relaxation times remain unknown, literature evidence suggests that cartilage hydration is likely dominant [34] [35]. Surprisingly, we observed higher T2 than T1ρ relaxation times in cartilage. In biological tissues, T1ρ relaxation times should be longer than T2 relaxation times because the spin-lock pulse forces the spins to precess about a direction different from the main magnetic field $B_0$, thereby slowing T2 relaxation [36]. Shorter repetition times (as present in the MIXTURE sequence) may have lead to T1ρ underestimation [37]: If the repetition time is too short, it may not allow for complete T1ρ relaxation and decrease T1ρ relaxation times. Other factors worth considering are the applied radiofrequency pulse for the T1ρ preparation, the $B_1$ inhomogeneity, and the magic angle effect [38]. Future phantom studies need to assess the accuracy and validity of MIXTURE-based relaxivity measurements versus reference measurements, e.g., multi-echo spin-echo sequences (for T2 quantification) and gradient-echo sequences (for T1ρ quantification) [39].

Our study has limitations. First, the in-situ defect model using human cadaveric knee joints only approximates the actual in-vivo situation. However, the model effectively excludes inter-sequence motion (and other artifacts such as arterial pulsations) and helps realize reproducible and standardized experimental conditions for voxel-wise comparisons. Regarding clinical translation, this model is inherently limited. Second, the number of specimens was small, and the study provided, by design, a focused proof of concept. Further diagnostic aspects relating to particular knee joint conditions require larger sample sizes and, ideally, assessment in the clinical routine.

In conclusion, combined morphologic and quantitative MRI sequences, such as the versatile MIXTURE platform, increase scanning efficiency and diagnostic utility by providing familiar contrasts and delivering additional quantitative information. In a basic research context, MIXTURE sequences demonstrated excellent delineability of cartilage defects and visualization of bone texture on par with the corresponding reference sequences. Once corroborated by larger clinical studies, MIXTURE may be a promising sequence platform for comprehensive and time-efficient joint imaging.

# References

[1] OARSI Osteoarthritis Research Society International, „Osteoarthritis: A Serious Disease, Submitted to the U.S. Food and Drug Administration December 1, 2016“.

[2] H. J. Braun und G. E. Gold, „Diagnosis of osteoarthritis: Imaging,“ *Bone,* Nr. 51, p. 278–288, 2012.

[3] R. Kijowski, D. G. Blankenbaker, d. M. A. Rio, G. S. Baer and B. K. Graf, "Evaluation of the Articular Cartilage of the Knee Joint: Value of Adding a T2 Mapping Sequence to a Routine MR Imaging Protocol," *Radiology,* no. 267(2), pp. 503-513, 2013.

[4] B. Hager, M. Raudner, V. Juras, O. Zaric, P. Szomolanyi, M. Schreiner und S. Trattnig, „MRI of Early OA,“ in *Early Osteoarthritis: State-of-the-Art Approaches to Diagnosis, Treatment and Controversies*, Cham, Springer, 2022, pp. 17-26.

[5] J. Le, Q. Peng und K. Sperling, „Biochemical magnetic resonance imaging of knee articular cartilage: T1rho and T2 mapping as cartilage degeneration biomarkers,“ *Annals of the New York Academy of Sciences,* Bd. 1383, Nr. 1, pp. 34-42, 2016.

[6] M. A. I. Alsayyad, K. A. A. Shehata und R. T. Khattab, „Role of adding T2 mapping sequence to the routine MR imaging protocol in the assessment of articular knee cartilage in osteoarthritis,“ *Egyptian Journal of Radiology and Nuclear Medicine,* Bd. 52, Nr. 78, 2021.

[7] H. F. Atkinson, T. B. Birmingham, R. F. Moyer, D. Yacoub, L. E. Kanko, D. M. Bryant, J. D. Thiessen und R. T. Thompson, „MRI T2 and T1ρ relaxation in patients at risk for knee osteoarthritis: a systematic review and meta-analysis,“ *BMC Musculoskelet Disord,* Bd. 20, p. 182, 2019.

[8] J. W. MacKay, S. B. L. Low, T. O. Smith, A. P. Toms, A. W. McCaskie und F. J. Gilbert, „Systematic review and meta-analysis of the reliability and discriminative validity of cartilage compositional MRI in knee osteoarthritis,“ *Osteoarthritis Cartilage,* Bd. 26, Nr. 9, pp. 1140-1152, 2018.

[9] T. M. Link, „Establishing compositional MRI of cartilage as a biomarker for clinical practice,“ *Osteoarthritis and Cartilage,* Bd. 26, Nr. 9, pp. 1137-1139, 2018.

[10] T. Hilbert, T. J. Sumpf, E. Weiland, J. Frahm, J.-P. Thiran, R. Meuli, T. Kober und G. Krueger, „Accelerated T2 Mapping Combining Parallel MRI and Model-Based Reconstruction: GRAPPATINI,“ *J Magn Reson Med,* Bd. 48, p. 359–368, 2018.

[11] S. M. Eijgenraam, A. S. Chaudhari, M. Reijman, S. M. A. Bierma-Zeinstra, B. A. Hargreaves, J. Runhaar, F. W. J. Heijboer, G. E. Gold und E. H. G. Oei, „Time-saving opportunities in knee osteoarthritis: T2 mapping and structural imaging of the knee using a single 5-min MRI scan,“ *European Radiology,* Bd. 30, p. 2231–2240, 2020.

[12] A. Chaudhari, K. Stevens, B. Sveinsson, J. Wood, C. Beaulieu, E. Oei, J. Rosenberg, F. Kogan, M. Alley, G. Gold und B. Hargreaves, „Combined 5-minute double-echo in steady-state with separated echoes and 2-minute proton-density-weighted 2D FSE sequence for comprehensive

whole-joint knee MRI assessment.," *J Magnetic Resonance Imaging,* Bd. 49, Nr. 7, pp. 183-194, 2019.

[13] A. S. Chaudhari, M. J. Grissom, Z. Fang, B. Sveinsson, J. H. Lee, G. E. Gold, B. A. Hargreaves und K. J. Stevens, „Diagnostic Accuracy of Quantitative Multicontrast 5-Minute Knee MRI Using Prospective Artificial Intelligence Image Quality Enhancement," *American Journal of Roentgenology,* Nr. 216, pp. 1-12, 2021.

[14] G. H. Welsch, K. Scheffler, T. C. Mamisch, T. Hughes, S. Millington, M. Deimling und a. S. Trattnig, „Rapid Estimation of Cartilage T2 Based on Double Echo at Steady State (DESS) With 3 Tesla," Bd. 62, p. 544–549, 2009.

[15] M. Yoneyama, T. Sakai, S. Zhang, D. Murayama, H. Yokota, Y. Zhao, S. Saruya, M. Suzuki, A. Watanabe, M. Niitsu und M. Van Cauteren, „MIXTURE: A novel sequence for simultaneous morphological and quantitative imaging based on multi-interleaved 3D turbo-spin echo MRI," in *Proc. ISMRM 2021:4203*, 2021.

[16] T. Sakai, M. Yoneyama, A. Watanabe, D. Murayama, S. Ochi, S. Zhang und T. Miyati, „Simultaneous anatomical, pathological and T2 quantitative knee imaging with 3D submillimeter isotropic resolution using MIXTURE," in *Proc. ISMRM 2021: 0845*, 2021.

[17] T. Sakai, J. Kwon, M. Yoneyama, D. Murayama, Q. Lu, A. Desai, A. S. Chaudhari, T. Miyati, S. Ochi und A. Watanabe, „MIXTURE-DOSMA: Initial clinical research of a comprehensive and multi-parametric quantitative 3D knee MR exam in patients with knee joint pain," in *Proc. Intl. Soc. Mag. Reson. Med. 30*, London, 2022.

[18] J. Kwon, T. Sakai, M. Yoneyama, D. Murayama, Q. Lu, A. D. Desai, A. S. Chaudhari und M. V. Cauteren, „MIXTURE-DOSMA: A comprehensive and multi-parametric quantitative 3D knee MR exam in 10 minutes," in *Proc. Intl. Soc. Mag. Reson. Med. 30*, London, 2022.

[19] D. Murayama, T. Sakai, M. Yoneyama und S. Ochi, „Simultaneous morphological and quantitative lumbar MRI with 3D isotropic high-resolution using MIXTURE T2," in *Proc. Intl. Soc. Mag. Reson. Med. 29*, Montreal, 2021.

[20] H. Yokota, T. Sakai, M. Yoneyama, Y. Zhao und T. Uno, „T2 Mapping of the Cranial Nerves with Multi-Interleaved X-prepared Turbo-spine Echo with Intuitive Relaxometry (MIXTURE) FLAIR," in *Proc. Intl. Soc. Mag. Reson. Med. 29*, Montreal, 2021.

[21] G. M. Sullivan und R. Feinn, „Using Effect Size—or Why the P Value Is Not Enough," *Journal of Graduate Medical Education,* Bd. 4, Nr. 3, pp. 279-282, 2012.

[22] "Python," [Online]. Available: https://www.python.org/downloads/release/python-399/. [Accessed 31 01 2023].

[23] "ITK-SNAP," [Online]. Available: www.itksnap.org. [Accessed 30 01 2023].

[24] P. A. Yushkevich, J. Piven, H. C. Hazlett, R. G. Smith, S. Ho, J. C. Gee und G. Gerig, „User-guided 3D active contour segmentation of anatomical structures: Significantly improved efficiency and reliability," *Neuroimage,* Bd. 31, Nr. 3, pp. 1116-28, 2006.

[25] M. Dieckmeyer, N. Sollmann, M. E. Husseini, A. Sekuboyina, M. T. Löffler, C. Zimmer, J. S. Kirschke, K. Subburaj und T. Baum, „Gender-, Age- and Region-Specific Characterization of

Vertebral Bone Microstructure Through Automated Segmentation and 3D Texture Analysis of Routine Abdominal CT,“ *Frontiers in Endocrinology,* Bd. 12, p. 792760, 2022.

[26] "pyRadiomics," [Online]. Available: https://pyradiomics.readthedocs.io/en/latest/. [Accessed 30 01 2023].

[27] "www.davuniversity.org," [Online]. Available: https://www.davuniversity.org/images/files/study-material/Texture%20feature_2-end.pdf. [Accessed 26 May 2023].

[28] A. C. o. Radiology, „ACR–SPR–SSR practice parameter for the performance and interpretation of magnetic resonance imaging (MRI) of the knee. Revised 2020 (Resolution 31),“ 2015.

[29] R. Kijowski, „3D MRI of Articular Cartilage,“ *Semin Musculoskelet Radiol,* Bd. 25, Nr. 3, pp. 397-408, 2021.

[30] O. Ristow, L. Steinbach, G. Sabo, R. Krug, M. Huber, I. Rauscher, B. Ma und T. M. Link, „Isotropic 3D fast spin-echo imaging versus standard 2D imaging at 3.0 T of the knee - image quality and diagnostic performance,“ *Eur Radiol,* Bd. 19, p. 1263–1272, 2009.

[31] J. P. Mugler, „Optimized Three-Dimensional Fast-Spin-Echo MRI,“ *Journal of Magnetic Resonance Imaging,* Bd. 39, p. 745–767, 2014.

[32] W. Wirth and F. Eckstein, "A technique for regional analysis of femorotibial cartilage thickness based on quantitative magnetic resonance imaging," *IEEE transactions on medical imaging,* no. 27(6), pp. 737-744, 2008.

[33] T. Nolte, S. Westfechtel, J. Schock, M. Knobe, T. Pastor, E. Pfaehler, C. Kuhl, D. Truhn und S. Nebelung, „Getting Cartilage Thickness Measurements Right: A Systematic Inter-Method Comparison Using MRI Data from the Osteoarthritis Initiative,“ *Cartilage,* 2023.

[34] K. Linka, M. Itskov, D. Truhn, S. Nebelung und J. Thuring, „T2 MR imaging vs. computational modeling of human articular cartilage tissue functionality,“ *Journal of the mechanical behavior of biomedical materials,* Bd. 74, pp. 477-487, 2017.

[35] J. Thuring, K. Linka, M. Itskov, M. Knobe, L. Hitpass, C. Kuhl, D. Truhn und S. Nebelung, „Multiparametric MRI and Computational Modelling in the Assessment of Human Articular Cartilage Properties: A Comprehensive Approach,“ *BioMed research international,* p. 9460456, 2018.

[36] Y.-X. J. Wáng, Q. Zhang, X. Li, W. Chen, A. Ahuja und J. Yuan, „T1ρ magnetic resonance: basic physics principles and applications in knee and intervertebral disc imaging,“ *Quantitative imaging in medicine and surgery,* Bd. 5, Nr. 6, pp. 858-885, 2015.

[37] W. L und R. RR, „T1rho MRI of human musculoskeletal system,“ *Journal of magnetic resonance imaging,* Bd. 41, Nr. 3, pp. 586-600, 2015.

[38] W. Chen, „Errors in quantitative T1rho imaging and the correction methods,“ *Quantitative Imaging in Medicine and Surgery,* Bd. 5, Nr. 4, pp. 583-591, 2015.

[39] N. A. Obuchowski, A. P. Reeves, E. P. Huang, X.-F. Wang, A. J. Buckler, H. J. (. Kim, H. X. Barnhart, E. F. Jackson, M. L. Giger, G. Pennello, A. Y. Toledano, J. Kalpathy-Cramer, T. V.

Apanasovich und e. al., „Quantitative imaging biomarkers: A review of statistical methods for computer algorithm comparisons," *Statistical Methods in Medical Research,* Bd. 24, Nr. 1, p. 68–106, 2015.

# Tables

## Table 1

**Table 1: MRI sequence parameters.** Two MIXTURE sequences were acquired, combining morphologic imaging with quantitative mapping, and 2D TSE and 3D TSE reference sequences of the same weighting. PD-w FS images were combined with quantitative T2 maps ("MIX 1") and T1-w images with T1ρ maps ("MIX 2"). Abbreviations: MIXTURE - **M**ulti-**I**nterleaved **X**-prepared **T**urbo-Spin Echo with Int**U**itive **Re**laxometry, PD - proton density, -w – weighted, FS - fat-saturated, TSE – turbo spin echo, TR – repetition time, TE – echo time, SENSE – sensitivity encoding, NSA – number of signal averages, SPAIR – spectral attenuated inversion recovery, SPIR – spectral presaturation with inversion recovery, N/A – not applicable, SL – spin lock, TSL – spin lock time, FOV – field of view, prep - preparation. Note that for 3D TSE sequences, $TE_{eff}$ and $TE_{equiv}$ denote the effective and equivalent TE as mediated by the choice of the refocusing pattern. In contrast, the 2D TSE sequence uses a constant refocusing flip angle that a single TE can describe. During the TSE readout, different refocusing patterns with variable order and magnitude of the flip angles are employed as designated by the manufacturer. T2-prep TE and SL-prep TSL refer to the duration of the preparation modules that MIXTURE employs to generate the respective contrast weightings.

| **Parameter** | **MIX 1** | **2D TSE PD-FS** | **3D TSE PD-FS** | **MIX 2** | **2D TSE T1-w** | **3D TSE T1-w** |
|---|---|---|---|---|---|---|
| Sequence type | 3D TSE | 2D TSE | 3D TSE | 3D TSE | 2D TSE | 3D TSE |
| Orientation | sagittal | | | | | |
| TR [ms] | 1200 | 3000 | 1100 | 600 | 582 | 400 |
| TE [ms] | N/A | 40 | N/A | N/A | 15 | N/A |
| $TE_{eff}$ [ms] | 125 | N/A | 125 | 22 | N/A | 36 |
| $TE_{equiv}$ [ms] | 46 | N/A | 46 | 13 | N/A | 21 |
| Echo train length [n] | 35 | 11 | 35 | 12 | 5 | 8 |
| Refocusing pattern | 'MSK PD FS' | 'no' | 'MSK PD FS' | 'Spine View T1' | 'constant' (110°) | 'MSK T1' |
| Compressed SENSE factor | 4.5 | 2.5 | 3.5 | 6 | 2 | 6 |
| NSA [n] | 1 | 2 | 1 | 1 | 2 | 2 |
| Fat saturation | SPAIR – none | SPIR | SPAIR | none – SPAIR | none | none |
| T2-prep TE [ms] | 0, 50 | N/A | N/A | N/A | N/A | N/A |
| SL-prep TSL [ms] | N/A | N/A | N/A | 0, 25, 50 | N/A | N/A |
| SL-prep frequency [Hz] | N/A | N/A | N/A | 500 | N/A | N/A |
| Scan time [min:s] | 4:59 | 4:06 | 2:57 | 6:38 | 4:23 | 4:22 |
| FOV [mm²] | 140 x 140 | | | | | |
| Acquisition matrix [px] | 304 × 304 | | | | | |
| Reconstruction matrix [px] | 512 × 512 | | | | | |
| Fat shift direction | Anteroposterior | | | | | |
| Phase Oversampling [%] | 12+12 | 30+30 | 12+12 | 12+12 | 33+33 | 12+12 |
| Slices [n] | 43 | | | | | |
| Slice thickness [mm] | 3 | | | | | |
| Slice oversampling [%] | 12 | N/A | 12 | 100 | N/A | 12 |

## Table 2

**Table 2: Quantification of Cartilage Defect Delineability.** The full widths at half maximum (FWHM) and edge widths (EW) were extracted from the line profiles of the PD-w FS images and used as surrogate parameters of defect delineability. EW was averaged over both defect shoulders. Data are presented as mean $\pm$ standard deviation [mm]. The statistical analysis was performed using repeated measures ANOVA. P-values are given as a function of sequence, delineability parameter (i.e., FWHM and EW), and nominal defect diameter (i.e., 3 mm, 5 mm, and 8 mm). Significant differences are indicated in **bold type**.

| **Nominal defect diameter** [mm] | **2D TSE**<br>FWHM [mm]<br>EW [mm] | **3D TSE**<br>FWHM [mm]<br>EW [mm] | **MIXTURE**<br>FWHM [mm]<br>EW [mm] | **P-value** |
|---|---|---|---|---|
| 3 | $2.6 \pm 0.4$<br>$1.0 \pm 0.4$ | $2.3 \pm 0.7$<br>$1.3 \pm 0.5$ | $2.6 \pm 0.3$<br>$1.3 \pm 0.6$ | p=0.34<br>p=0.02 |
| 5 | $4.4 \pm 0.2$<br>$1.1 \pm 0.7$ | $4.3 \pm 0.2$<br>$1.4 \pm 0.4$ | $4.3 \pm 0.2$<br>$1.4 \pm 0.4$ | **P=0.004†**<br>p=0.04 |
| 8 | $7.1 \pm 0.5$<br>$1.5 \pm 0.9$ | $7.0 \pm 0.5$<br>$1.8 \pm 1.1$ | $7.0 \pm 0.5$<br>$1.8 \pm 1.0$ | p=0.47<br>p=0.45 |

(†) The posthoc details (Tukey's test) regarding multiplicity-adjusted p-values for pairwise sequence comparisons were p=0.03 for 2D TSE vs. 3D TSE, **p=0.005** for 2D TSE vs. MIXTURE, and p=0.91 for 3D TSE vs. MIXTURE.

## Table 3

**Table 3: Quantification of cartilage composition and ultrastructure.** T2 and T1ρ relaxation times of the segmented cartilage of the central lateral femorotibial compartment before and after defect creation in ten knee joint specimens. Mean ± standard deviation [ms]. The regional assessment included three femoral and one tibial region. Pre-defect and post-defect relaxation times were compared using the Wilcoxon matched-pairs signed rank test, and multiplicity-adjusted p-values were determined. Significant differences are indicated in **bold type**.

| **Region** | **$T2^{pre}$ [ms]** | **$T2^{post}$ [ms]** | **P-value** | **$T1\rho^{pre}$ [ms]** | **$T1\rho^{post}$ [ms]** | **P-value** |
|---|---|---|---|---|---|---|
| Anterior Femur | $48 \pm 3$ | $49 \pm 3$ | P=0.43 | $43 \pm 5$ | $46 \pm 3$ | P=0.04 |
| Central Femur | $51 \pm 4$ | $56 \pm 4$ | **P=0.002** | $41 \pm 5$ | $43 \pm 5$ | P=0.19 |
| Posterior Femur | $64 \pm 8$ | $69 \pm 9$ | P=0.04 | $34 \pm 9$ | $39 \pm 8$ | P=0.02 |
| Tibia | $41 \pm 4$ | $45 \pm 4$ | P=0.05 | $36 \pm 4$ | $40 \pm 5$ | P=0.01 |
| All regions | $48 \pm 2$ | $51 \pm 2$ | P=0.03 | $40 \pm 4$ | $43 \pm 4$ | **P=0.004** |

# Figures

## Figure 1

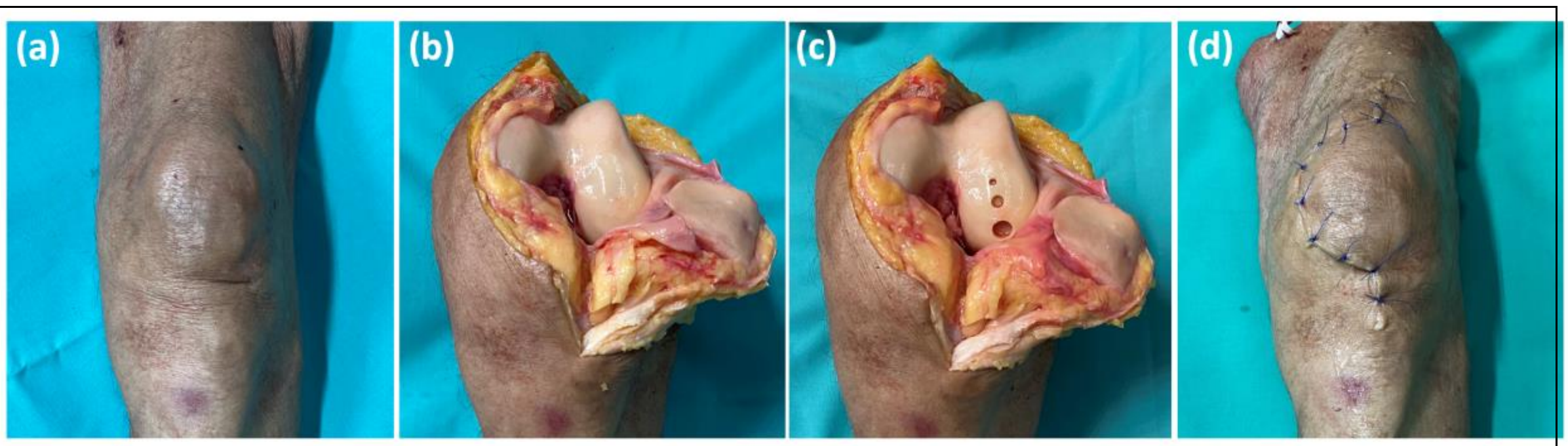

**Figure 1: Standardized Cartilage Defect Model.** (a) Intact knee joint. (b) Complete surgical exposure of the knee joint specimen through the longitudinal arthrotomy, medial peri-patellar incision, and lateral eversion of the patella. (c) By use of biopsy punches, cartilage defects of variable diameters, i.e., 3 mm (top), 5 mm (center), and 8 mm (bottom), were aligned anteroposteriorly. (d) The wound was closed by layer-wise suturing under continuous irrigation.

## Figure 2

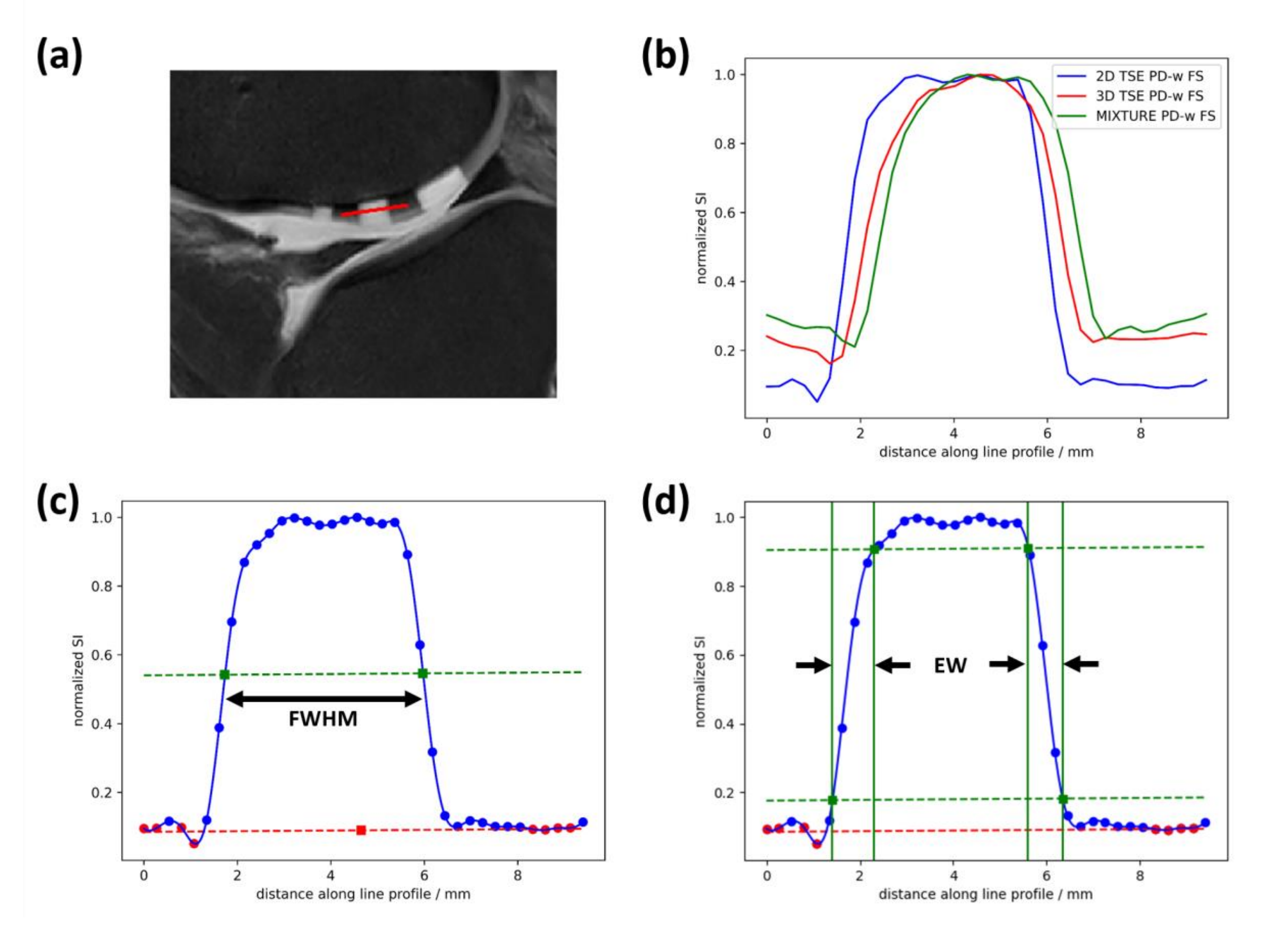

**Figure 2: Inter-sequence comparison of cartilage defect delineability.** (a) For this representative defect of 5 mm diameter, a line was manually annotated to transect the defect and adjacent cartilage at mid-substance (red line, sagittal PD-w FS image). (b) For each sequence, i.e., 2D TSE, 3D TSE, and MIXTURE, the line profiles (corresponding to the pixel-wise signal intensity along the red line) were extracted, normalized to the maximum signal intensity of 1 (blue circles) and used to calculate the full width at half maximum (FWHM, c) and the edge widths (EW, d) as surrogate measures of defect delineability. (c) FWHM was determined by determining the half maximum (dashed green line) between the cartilage background signal intensity (dashed red line) and the maximum signal intensity and by measuring the horizontal distance between the intersecting points of the half maximum with the signal intensity profile (green dots). (d) Analogously, EW was determined by defining the horizontal distances between the 10% and 90% maximum intensity levels (green dots, dashed green lines) on both defect shoulders.

## Figure 3

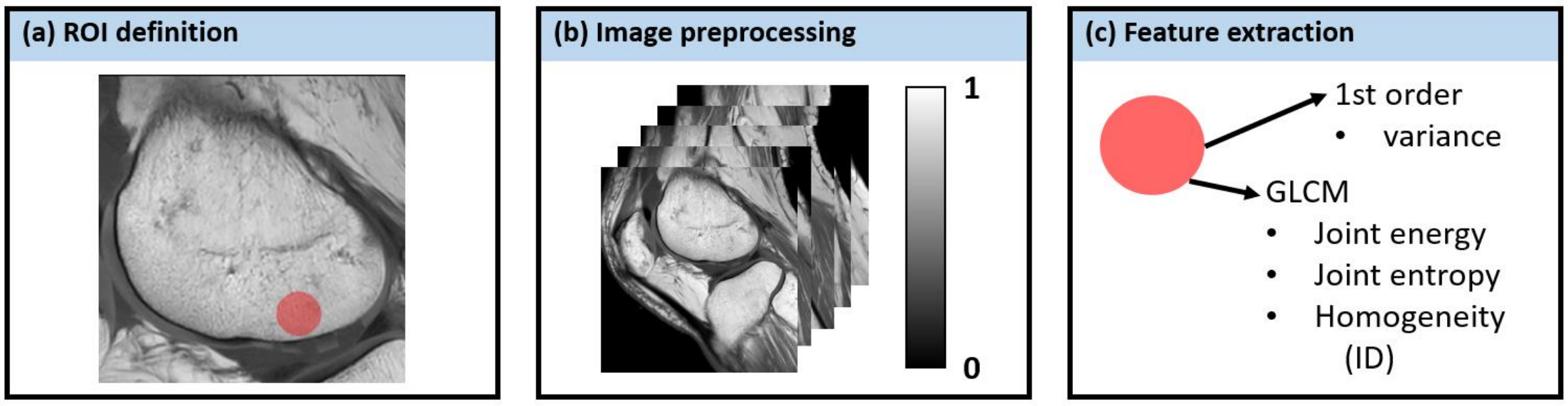

**Figure 3: Inter-sequence comparison of bone texture.** On the T1-weighted images, radiomic features were analyzed by defining a standardized circular region of interest adjacent to the subchondral lamella underneath the 5 mm defect (a). Image preprocessing included normalization of the signal intensity to the range of 0 to 1 (b). As texture features, variance as a first-order feature and joint energy, joint entropy, and homogeneity (i.e., the inverse difference [ID]) as gray level co-occurrence matrix (GLCM) features were extracted (c). For the computation of the GLCM, the normalized images were quantized into 200 evenly spaced bins.

## Figure 4

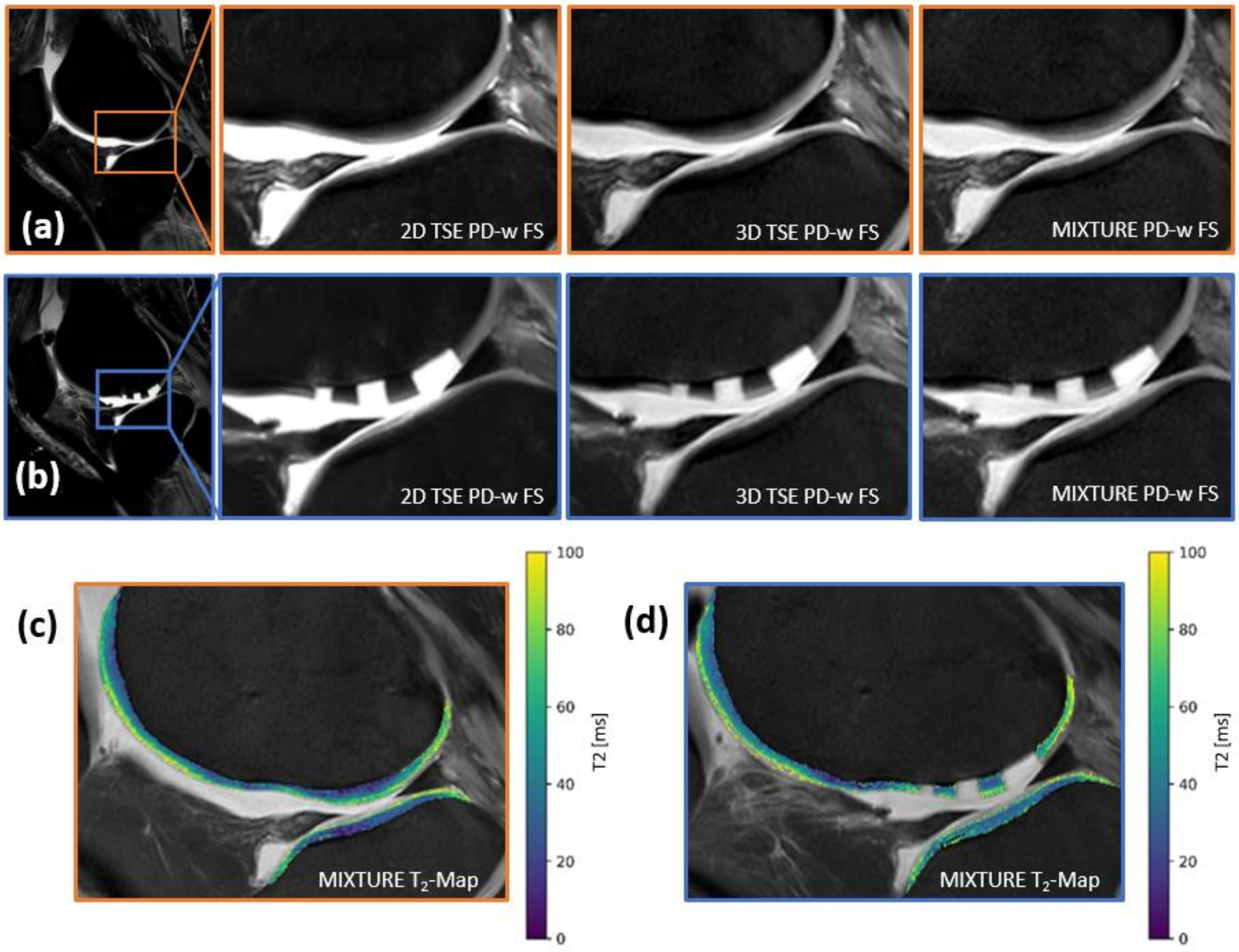

**Figure 4: Representative PD-weighted fat-saturated images and MIXTURE T2 maps.** Sagittal images before (orange frame, a) and after (blue frame, b) the creation of standardized cartilage defects. The slice that centrally bisected the three defects and the corresponding slice of the intact joint was selected. Cartilage defects of 3 mm, 5 mm, and 8 mm diameter (from left [anterior] to right [posterior]) are displayed. Zoomed images (indicated by the inset boxes in the leftmost images) are from left to right: the 2D TSE sequence, the 3D TSE sequence, and the MIXTURE sequence. Corresponding MIXTURE-based T2 maps before (c) and after (d) defect creation. Scale bar on the right extends from 0 to 100 ms.

## Figure 5

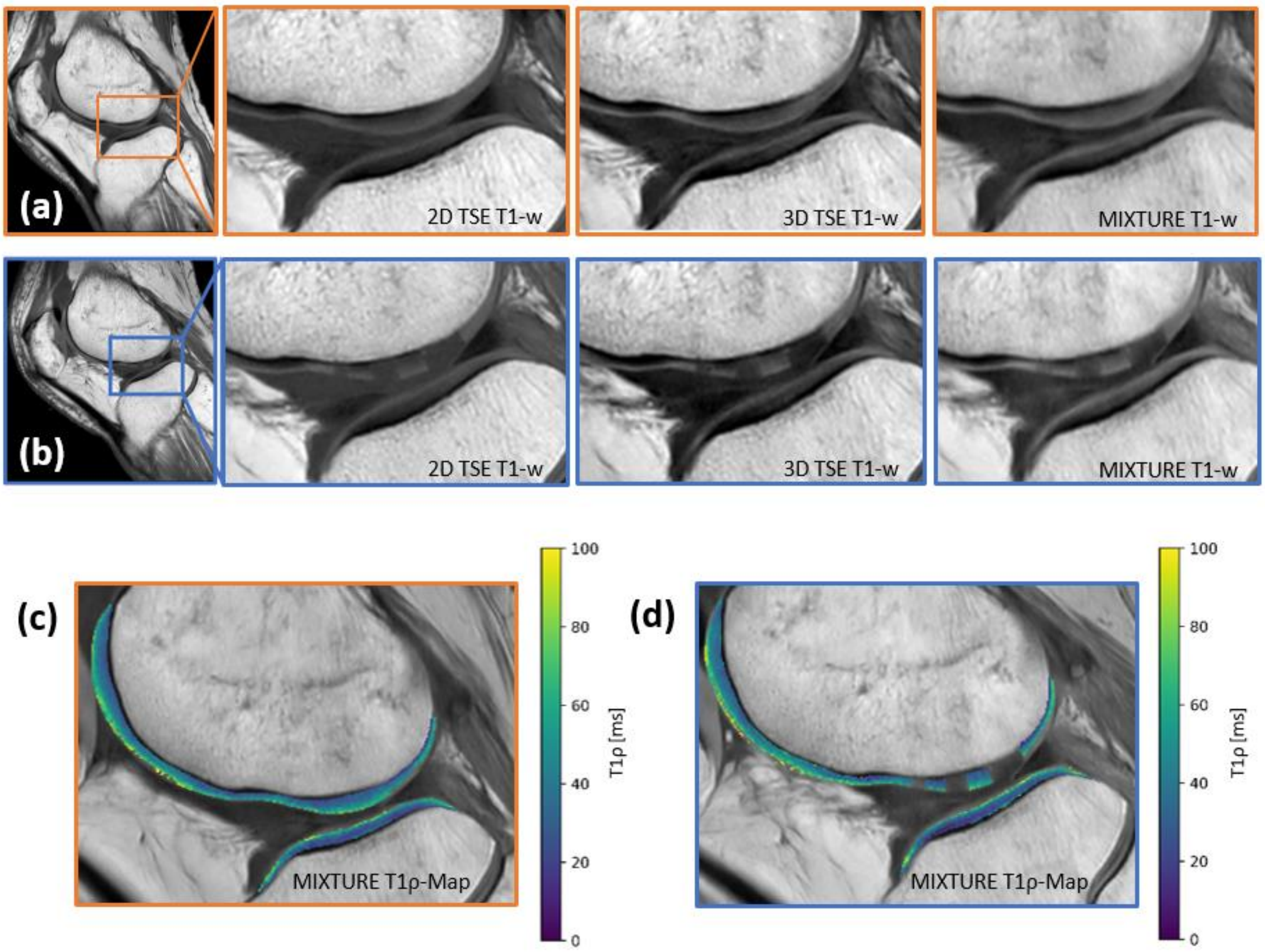

**Figure 5: Representative T1-weighted images and MIXTURE T1ρ maps.** Morphologic images (a, b) and corresponding MIXTURE-based T1ρ maps (c, d) are visualized before and after defect creation. Figure organization as in **Figure 4**.

# Figure 6

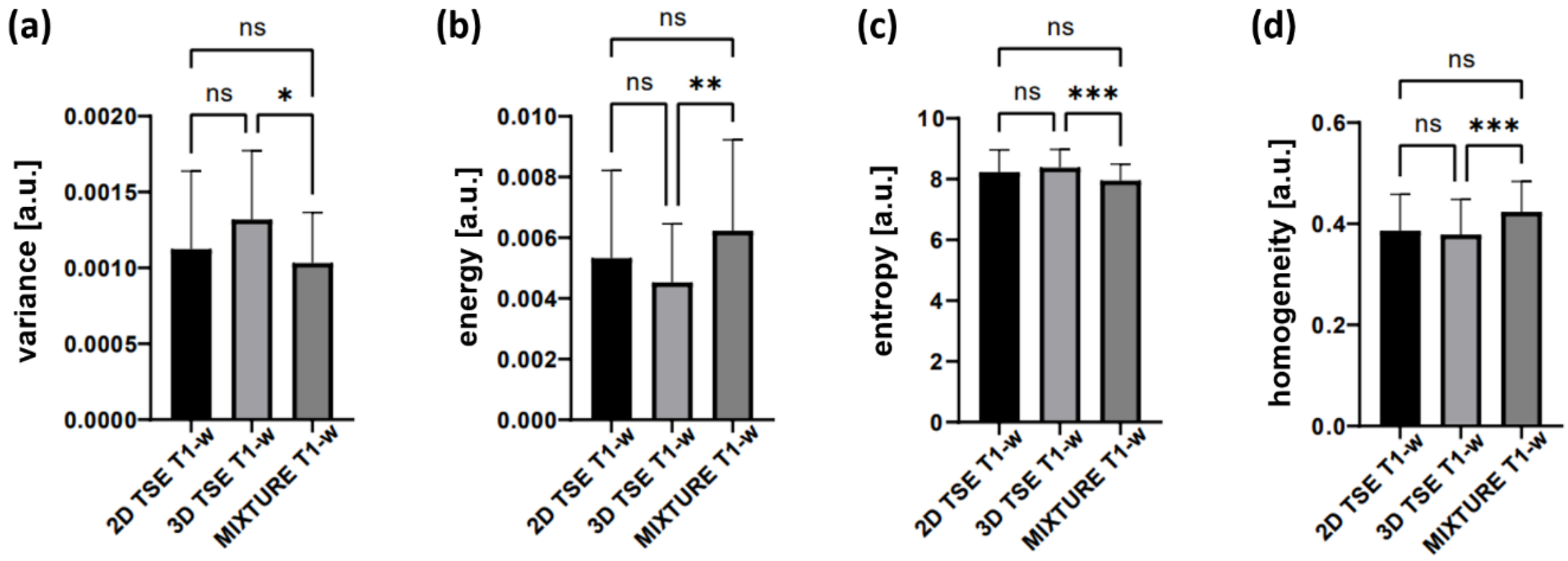

**Figure 6: Analysis of bone texture features.** Based on the radiomic feature analysis, a representative region of interest of the subchondral bone was defined and compared between the T1-weighted sequences, i.e., the 2D TSE, 3D TSE, and MIXTURE sequences. Variance (a), energy (b), entropy (c), and homoegeneity (d) were quantified and analyzed as measures of bone texture. Levels of statistical significance were stratified as 'ns', '*', '**', and '***' to indicate $p>0.05$, $0.01<p\leq0.05$, $0.001<p\leq0.01$, and $p\leq0.001$, respectively.